\documentclass[11pt]{article}

\usepackage{amsmath,amssymb}
\usepackage{graphicx}
\usepackage{epstopdf}
\newcommand{\be}{\begin{equation}}
\newcommand{\ee}{\end{equation}}
\newcommand{\bea}{\begin{eqnarray}}
\newcommand{\eea}{\end{eqnarray}}
\newcommand{\ba}{\begin{array}}
\newcommand{\ea}{\end{array}}

\newcommand{\tr}{\mbox{tr}}

\newcommand{\la}{\langle}
\newcommand{\ra}{\rangle}

\newcommand{\cG}{{\cal{G}}}

\numberwithin{equation}{section}

\newcommand{\da}{{\dot{a}}}

\newcommand{\bra}[1]{\left\langle \left. #1 \right| \right.}
\newcommand{\ket}[1]{\left| \left. #1 \right\rangle \right.}

\newcommand{\nn}{\nonumber}

\begin{document}

\allowdisplaybreaks

\title{
	Emission spectrum of soft massless states from heavy superstring \bigskip}

\author{
        Shoichi Kawamoto$^*$ 
        and
	Toshihiro Matsuo$^\dagger$\bigskip
	\\
	$^*$\footnotesize\it Department of Physics, Tunghai University,
Taichung 40704, Taiwan\\
	\footnotesize\tt kawamoto@thu.edu.tw, kawamoto@yukawa.kyoto-u.ac.jp
	\smallskip\\
	$^\dagger$\footnotesize\it Anan National College of Technology, Tokushima 774-0017, Japan\\
	\footnotesize\tt tmatsuo@yukawa.kyoto-u.ac.jp
	}

\date{\today}

\maketitle

\bigskip

\begin{abstract}
\noindent\normalsize
We calculate emission rates of various bosonic/fermionic soft massless states
of an open/closed superstring from an ensemble of a
highly excited open/closed superstring in the flat background.
The resulting spectrum shows thermal distributions at the Hagedorn
temperature.
We find greybody factors for each process
and observe their relation to the ones from  black holes.
\end{abstract}

\vfill

\setcounter{footnote}{0}
\section{Introduction}

Highly excited states of string theory are 
of particular interest in perturbative string theory.
An exponential growing number of states at higher levels
leads to a characteristic temperature of the string ensemble,
the Hagedorn temperature, at which the partition function gets
divergent.
This divergence may be interpreted
as a signal of a 
phase transition;
above this temperature, string theory has been speculated to have
much fewer degrees of freedom than any kind of quantum field theory\cite{Atick:1988si}.
This would be related to rich ``stringy symmetries''
that might emerge at a scale much higher than the string scale\cite{Gross:1988ue}.
Furthermore, in various extreme situations,
such as an early Universe or high-energy scattering processes,
highly excited states can be created and then their properties
would be important to applications of string theory\footnote{%
  An example of recent application is found in \cite{Skliros:2013pka}}.

Excited states of a string are usually unstable and decay eventually.
There have been lots of studies on this instability,
such as a typical lifetime or a decay spectrum\cite{splitting_prob,decays}.
One of the interesting setups to investigate this is
a semi-inclusive decay process, where only the mass (and the angular momentum
in some cases) of the initial state is fixed.
By taking an average over the initial states, the process exhibits
a thermodynamic behavior.
Amati and Russo\cite{Amati:1999fv}
have shown that the decay spectrum of the highly excited fundamental bosonic string
is the thermal one of the Hagedorn temperature.
Since then, there have been many works on this type of
 analysis
for boson emission of an NS-R superstring and also closed string
states emission from a heavy closed string\cite{Manes:2001cs, Chen:2005ra},
 especially on the decay rate
of maximally angular momentum states with interest in searching for
 possible long-lived states\cite{Decay_w_ang_mom}.
Some other applications of this procedure for the cross section
of strings are found in \cite{Kuroki:2007aj, Matsuo:2009sx}.

Another motivation for the study on the decay of a heavy string
 is regarding black hole physics.
A couple of decades ago, Susskind\cite{Susskind:1993ws} proposed that
the microstates of a black hole could be explained
by an exponentially growing number of states of a heavy fundamental string.
This correspondence is considered to
take place at a point where the typical size of a free string
of a given mass becomes a size of the Schwarzshild radius with respect to that mass.
The entropy of these two descriptions becomes the same order at that point.
This idea was pursued further by Horowitz and Polchinski\cite{Horowitz:1996nw},
and they showed that
this correspondence indeed holds for various types of black holes.
The corresponding point of a black hole and a fundamental string
typically appears at $g_s\sim N^{-1/4}$, where $g_s$ is the string coupling
constant and $N$ is the excitation level of the fundamental string.
For very large $N$, the leading-order treatment of this heavy string
by perturbation theory would work, and
may capture some aspects of the corresponding black holes\cite{Matsuo:2008fj}.
Among others, one of the characteristics of a black hole is its greybody factor.
Although black holes exhibit the blackbody radiation at the horizon,
their gravitational potentials alter the spectrum for an asymptotic observer.
This correction factor, known as the greybody factor, was actually
an important clue for string/gauge theory correspondence in the early days of
its development\cite{Das:1996wn, Maldacena:1996ix}.
The greybody factor for a near-BPS black hole that has D-brane construction
shows a perfect agreement with the calculation of gauge theory on the branes.
As we will show in this paper, the decay rates of a heavy superstring indeed
 turn out to exhibit a thermal behavior of the Hagedorn temperature
and we can read off corresponding greybody factors for the heavy superstring.
We may, thus, expect that similar insight can be obtained for a more general class of 
black holes through the study of fundamental string decay.
It should be noted that, as explained in the main part of the paper,
our analysis is to the leading order of perturbation theory as well as
to the leading order of $1/N$ expansion, and 
the correspondence point is not reached completely
within this regime.
For example, if we want to obtain the spectrum of the Hawking temperature, 
instead of that of the Hagedorn temperature, we would need to take the self-gravitational effect
into account\cite{Horowitz:1997jc, Damour:1999aw},
but we will neglect self-interactions in this paper.
However, we believe that the current study can be thought of as a first step
toward this understanding, and it deserves more detailed study in the future.

In this paper, we consider 
single open/closed string massless state emission from
the decay of a massive open/closed superstring in the critical
dimensions.
As an initial state, we prepare an averaged open or closed superstring state
at a very high excited level.
We only specify the mass (and, therefore, the excited level) of the string,
and also observe the energy spectrum of the emitted states.
As the emitted states, we consider both open and closed string states.
In the perturbative regime, we can take the massless states as the main channel of decay.
We also integrate over the angular dependence and sum over
the polarizations of the emitted massless states.
We will work with Green-Schwarz formulation of superstrings in the light-cone gauge.
It has an advantage that the physical degrees of freedom are explicit, and
we do not need to worry about the treatment of unphysical modes.
As we shall see, our setup is fit to the restriction of the momentum of the vertex operators
in the light-cone gauge, and we can carry out the whole calculation very explicitly.

The organization of this paper is as follows.
In Section \ref{sec:emiss-massl-stat}, 
we shall present a setup of a semi-inclusive decay process of a heavy superstring.
Then, we carry out the calculation of the emission rates of massless
states from a heavy open superstring, by use of Green-Schwarz
formalism
in the light-cone gauge.
We also argue the closed string emission from 
both heavy open and closed superstrings.
We conclude this section with a detailed argument of the emission rates
of each case.
There, we compare the greybody factors obtained from the emission rates
with the ones from various types of black holes.
In Section \ref{sec:conlusion-discussion},
we summarize our result and propose possible future directions.
Appendix \ref{sec:misc-calc} is devoted to the summary of the details
of the calculation.

\section{Emission of massless states}
\label{sec:emiss-massl-stat}

\subsection{Semi-inclusive decay process}
\label{sec:incl-decay-proc}

As stated in the introduction,
we observe emission from a heavy superstring at an asymptotic infinity
and will be ignorant about the detailed profile of the initial and final string states.
We shall study a semi-inclusive decay process of a highly excited
superstring in the critical dimensions with a massless state (either bosonic or fermionic) emitted.
The emitted massless state is characterized by its momentum $k^\mu$
and polarization tensor $\gamma(k)$.
The initial state is at an excited level $N$ and carries momentum $P^\mu_\text{ini}$.
It decays into a state at level $N'$ with $P^{\mu}_\text{fin}$
with a massless state emitted. 

First, we choose the center-of-mass frame of the initial string,
$P_\text{ini}^\mu=(M, \vec{0})$,
with $\sqrt{\alpha'} M =\mathcal{O}(\sqrt{N})$.
In this frame, the momentum
of the emitted state is to be $k^\mu = (-\omega, \vec{k})$
with $\omega^2=|\vec{k}|^2$ as it is massless, and then by momentum conservation,
the final-state momentum is determined as
$P_\text{fin}^\mu=(-M+\omega, -\vec{k})$.
The (differential) decay rate is given by
\begin{align}
  \Gamma =& \frac{d^{9} k}{M(M-\omega)\omega} P(\Phi_N \rightarrow \gamma(k) +
  \Phi_{N'}) \,,
\end{align}
where $\Phi_N$ denotes arbitrary states of string at the level $N$,
and the probability $P(\Phi_N \rightarrow \gamma(k) +
  \Phi_{N'})$ is the modulus square of the amplitude of the process.
We will not be interested in the angular dependence of emission,
and $d^9 k$ will eventually be set to $\omega^8 d \omega$.
The masses of both the initial and the final string are heavy and 
are assumed to be much larger
than the typical energy of the emitted massless states, $M \gg \omega$.

We are considering the semi-inclusive decay process.
We specify only the mass (therefore, the level) of the final state
and are also interested in the energy distribution of the emitted states.
We do not consider all possible final states, which may involve
multistring states and many light states, but rather restrict ourselves to
this three-body decay process; namely, we are working in the leading order
of perturbation theory for a given process.
In summary, for calculations of probability, we sum over all possible
states of the final string states $\Phi(N')$ and the emitted massless
state
$\gamma(k)$, as well as the angular part of $k^\mu$.
As for the initial state,
we do not prepare any particular state of mass $M$
but rather take a typical state by averaging over the possible states of the initial string
at a given level.
The probability is, thus,
\bea
P(\Phi_{N} \to \gamma(k)+\Phi_{N'})
=\frac{1}{\cG(N)}\sum_{\Phi|N}\sum_{\Phi|N'}\sum_{\gamma{}}
|\la \Phi(N')|V(\gamma{}, k)|\Phi(N)\ra|^2 ,
\eea
where
$\sum_{\Phi|N}$ represents the summation over all the states at level $N$,
and $\Phi(N)$ stands for a state at level $N$.
The number of states at level $N$ is denoted by $\cG(N)$, and 
the asymptotic form of $\cG(N)$ at large $N$ is calculated 
in Appendix \ref{sec:dens-stat-haged}.
$V(\gamma,k)$ is the string vertex operator corresponding to the emitted state.

In general, it is a formidable task to handle a general string state at a high
fixed level, due to the exponentially growing number of states.
We trim the expression of the probability
to make it a more tractable form, following the trick initiated by \cite{Amati:1999fv}.
It is convenient to introduce a projection operator onto the level $N$ states
\bea
\hat{P}_N = \oint \frac{dv}{2\pi i v} v^{\hat{N}-N} \,,
\qquad
\sum_{\Phi|N} \ket{\Phi} = 
\sum_{\Phi} \hat{P}_N \ket{\Phi}
\,,
\label{projectionP}
\eea 
where the sum in the right-hand side of the second equation runs over all the states
in Fock space.
Then, the probability is written as
\begin{align}
P(\Phi_{N} \to \gamma(k)+\Phi_{N'})
=&
\frac{1}{\cG(N)}\sum_{\gamma}
\sum_{\Phi,\Phi'}
\big| \bra{\Phi'} \hat{P}_{N'}\,  V(\gamma,k) \,  \hat{P}_N \ket{\Phi} \big|^2
\nn\\ =&
\frac{1}{\cG(N)}\sum_{\gamma}
\oint \frac{dw}{2\pi i w}w^{-N}
\oint \frac{dv}{2\pi i v}v^{-N'}
\tr[V^\dag(\gamma, k)\, v^{\hat{N}}\, V(\gamma, k)\, w^{\hat{N}}] 
\nn\\ =&
\frac{1}{\cG(N)}\sum_{\gamma}
\oint \frac{dw}{2\pi i w}w^{-N}
\oint \frac{dv}{2\pi i v}v^{N-N'}
\tr[V(\gamma,k,1)^\dagger\,  V(\gamma, k,v) \, w^{\hat{N}}] 
\,,
  \label{eq:5}
\end{align}
where the trace is taken in Fock space, namely, for the oscillator part.
In the last line, we have used the fact that the operator
$v^{\hat{N}}$ transports the (oscillator part of the)
vertex  operator to the position $v$ as
$v^{\hat{N}}V(\gamma, k, 1)v^{-\hat{N}}=V(\gamma, k, v)$.
The third entry of the vertex operator now stands
 for the insertion point along world-sheet time direction $\tau$ with $v=e^{i\tau}$.
As for the bosonic zero-mode part,
the momentum operators will be evaluated as
the initial- or final-state momentum value since this is a disk amplitude.
The other contribution from the bosonic zero modes is a trivial momentum conservation factor
that we do not write down explicitly in this paper.
The trace part appears as a similar form to the oscillator part of
string one-loop computations.
However, it should be noted that there is a crucial difference;
the trace here is originated from the square of the disk
amplitude and, thus,
not the supertrace defined with the $(-1)^F$ operator inserted.
Therefore, it is different from superstring one-loop amplitudes,
and the result is nonvanishing even though we have only two vertex operators inserted.
This \eqref{eq:5} 
is the master formula for the semi-inclusive decay process we are going to study.
We will evaluate this trace in the open and closed superstring theory,
with the identical two vertex operators for both open and closed massless states
inserted.

A couple of comments on the emission of other states are in order.
A heavy string can emit massive states
or split into heavy strings, too.
Now, we briefly argue that the emission of soft massless states
is a dominant channel of decay.

A heavy string may split into two heavy strings.
In this case, two final states have string scale masses,
$M^2 \sim \mathcal{O}(N)/\alpha'$.
Starting from the rest frame of the initial string,
these two strings move much more slowly than light states
unless their spatial momenta
are of $\mathcal{O}(N)$ in the string scale.
Therefore, when we consider ourselves as an asymptotic observer,
we would have little chance to detect such heavy string states.
Note that once higher-order effects are included, these kinds of end states
are more irrelevant, as they are bound by their own gravitational potential.
In the exponentially many number of possible states, noninteractive pairs,
like BPS configurations, would be negligibly scarce.
A heavy state may further decay into lighter states and eventually
emit sufficiently light states that can propagate far enough to be detected.
There can be enormous intermediate steps to end up with light states,
and then
these processes may be favored with respect to an entropic viewpoint.
However, in this paper, we consider only the leading-order contribution
of string perturbation theory and will not take this multistep
decay process into account.
It is interesting to investigate the competition between
the growing number of possible intermediate states and the suppressing power of coupling
constant, but it is beyond the scope of our current study.

Finally, we consider the contribution from rather light but
stringy massive states.
Since we study  highly excited string states, then these lowest-level states
may be regarded as light states to enter our consideration.
As we will see, it turns out that the emission spectrum for massless states becomes
a thermal one at the Hagedorn temperature.
The Hagedorn temperature of the superstring, $T_H=(\pi \sqrt{8\alpha'})^{-1}$,
is numerically smaller than the mass of the first excited state,
$M_1 =c (\alpha')^{-1/2}$, where $c=1$ for open and $c=2$ for closed string states.
Therefore, in the thermal distribution of the Hagedorn temperature,
the massive states will hardly be observed,
and we concentrate on massless states emissions.
From the same reason, the energy of the emitted massless state should also be
much smaller than the string scale.
Therefore, we take the emission of soft massless states 
as the main channel of the decay process in this paper.

\subsection{Open string emission from an open superstring}
\label{sec:open-string-emission}

First, we consider an open string emission rate from 
a heavy open superstring.
In this case, the mass of the initial and the final states
are $M=\sqrt{N/\alpha'}$ and $M'=\sqrt{N'/\alpha'}$.
From the momentum conservation, we find the level difference
between the initial and the final state is $\mathcal{O}(\sqrt{N})$,
\begin{align}
  \label{eq:80}
  N - N' =& 2\omega \sqrt{\alpha' N} + \alpha' \omega^2 \,,
\end{align}
and the last term is negligible as $\sqrt{\alpha'}\omega \ll \sqrt{N}$.
We now explicitly evaluate the traces of
massless boson and fermion vertex operators
shown in the previous section.
From now on, we set the Regge slope parameter $\alpha'=1/2$
for simplicity.

In Green-Schwarz superstring in the light-cone gauge, 
the vertex operators for massless boson and fermion states are
\begin{align}
  \label{eq:7}
  V_B(\zeta,k,z) =& \left(
\zeta^i(k) B^i - \zeta^-(k) {p}^+
\right) e^{i k \cdot X(z)} \,,
\\
  V_F(u,k,z) =& \left(
u^a(k) F^a + u^\da(k) F^\da
\right) e^{i k \cdot X(z)} \,,
\end{align}
where $B^i$, $F^a$, and $F^\da$ are represented by the light-cone
fields\cite{GSW}.
The explicit forms are given in Appendix \ref{sec:calculation-trace}.
It should be noted that these vertex operators are valid only for
the emission with momentum $k^+=0$,
and otherwise they take more complicated forms.
Since we have neglected the angular distribution of the momentum of
the emitted states, we can choose the momentum
$k^\mu = (-\omega,0,\cdots,0,\omega)$ by transverse
$SO(9)$ rotation of the rest
frame of the initial string.
So we can consistently choose the light-cone coordinate so that $k^+=0$ for
the emitted state.

We are going to calculate the decay rate of a heavy open superstring
with a massless boson/fermion state emitted,
\begin{align}
  \label{eq:15}
  \Gamma_A =& \frac{\omega^7 d\omega}{M^2} 
P\big(\Phi_N \rightarrow \gamma_A(k) + \Phi_{N'} \big)
\,,
\end{align}
where $P\big(\Phi_N \rightarrow \gamma_A(k) + \Phi_{N'} \big)$ is
given by \eqref{eq:5} with the vertex operator $V_A$,
where  $A=B$ for boson emission and $F$ for fermion emission.
In the same way,
the polarization is $\gamma_B = \zeta^i, \zeta^-$ or $\gamma_F=u^a, u^\da$.
What we need to do first is to calculate the oscillator trace
and then evaluate the $v$ and $w$ integral to derive
the probability $P$.
The explicit calculation is straightforward but rather lengthy.
It is summarized in Appendix \ref{sec:open-string-vertex}.
We cite the final result of the trace calculation,
\begin{align}
  \label{eq:8}
  \tr\left(
V_B(\zeta,k,1)^\dagger V_B(\zeta,k,v) \, w^{\hat{N}}
\right) =&
\left(
|\zeta^i|^2 \Omega(v,w)
+|\zeta^-|^2 (P_\text{ini}^+)^2
\right)
Z(w)
\,,\\
  \tr\left(
V_F(u,k,1)^\dagger V_F(u,k,v) \, w^{\hat{N}}
\right) =&
\frac{1}{4} \bigg[  P_\text{ini}^+
u^{a *} u^a
+ u^{\dot a *} \gamma_{\dot a b}^i u^{ b} P_\text{ini}^i
+ u^{ a *} \gamma_{a\dot b}^i u^{\dot b} P_\text{ini}^i
\nn\\&\hskip2em
+  \frac{ u^{\dot a *} u^{\dot a}}{P_\text{ini}^+}
\big((P_\text{ini}^i)^2+  \Omega(v,w)  \big)
\bigg]
\Xi(v,w)
Z(w)
 \,,
\end{align}
where
\begin{align}
  \label{eq:9}
\Omega(v,w) =
\sum_{n=1}^\infty  \,
n \frac{v^n+(w/v)^n}{1-w^n}
\,,
\qquad
\Xi(v,w)=
\frac{1}{2}+
 \sum_{n=1}^\infty
\frac{v^n + (w/v)^n}{1+w^n}
\,,
\end{align}
and $Z(w)$ is the partition function,
\begin{align}
  \label{eq:68}
Z(w)=& 16 \, \left(\frac{f_+(w)}{f_-(w)} \right)^8
\,, \qquad
    {f}_\pm (w)=
\prod_{n=1}^\infty (1\pm w^n)
\,.
\end{align}
$f_\pm(w)$ are contributions from bosonic oscillators ($-$) and fermionic ones ($+$),
respectively. 
The value of $16$ is from the vacuum degeneracy due to the fermionic zero modes.
As noted in the introduction, the trace here is different from the supertrace of the superstring
one-loop calculation, and then the bosonic and fermionic parts do not cancel and lead to
the partition function.
Its asymptotic behavior is evaluated in Appendix \ref{sec:dens-stat-haged}.
Since the initial momentum is given by $P_\text{ini}^i=0$ and $P_\text{ini}^+=\sqrt{N}$, the terms multiplied by $P_\text{ini}^i$ vanish in the expression.

Let us start with the boson emission process:
\begin{align}
  \label{eq:1}
P(\Phi_{N} \to \zeta(k)+\Phi_{N'})
=
\frac{1}{\cG(N)}\sum_{\zeta}
\oint \frac{dw}{2\pi i w}w^{-N}
\oint \frac{dv}{2\pi i v}v^{N-N'}\left(
|\zeta^i|^2 \Omega(v,w)
+N |\zeta^-|^2 
\right)
Z(w)
\,.
\end{align}
After the contour integration with respect to $v$, only the term
$v^{-n}$ with $n=N-N'$ in the $\Omega$ survives.
Note that $N>N'$.
Thus, we have 
\begin{align}
  \label{eq:10}
P(\Phi_{N} \to \zeta(k)+\Phi_{N'})
=&
\frac{1}{\cG(N)}\sum_{\zeta} |\zeta^i|^2 
\oint \frac{dw}{2\pi i w}
\frac{(N-N')w^{-N'}}{1-w^{N-N'}}
Z(w)
\,.
\end{align}
For large $N'$, the integral can be evaluated by the saddle point
method.
For $w=e^{-\beta}$, the dominant contribution will come
from $\beta\simeq 0$.
By using the modular transformation property of the partition function, which is shown 
in Appendix \ref{sec:dens-stat-haged},
one finds a saddle point at $\beta = {\pi \sqrt{2/N'}}$.
After the Gaussian integration around the saddle point and noting $\sqrt{N}-\sqrt{N'} \simeq \omega$,
 we obtain
\begin{align}
  \label{eq:11}
P(\Phi_{N} \to \zeta(k)+\Phi_{N'})
\simeq &
\frac{1}{\mathcal{G}(N)} \sum_{\zeta}
|\zeta^i|^2
\frac{(N-N') e^{\pi\sqrt{8N'}}N^{\prime
    -\frac{11}{4}}}{1-e^{-\sqrt{2}\pi\frac{N-N'}{\sqrt{N'}}}}
\big(1+\mathcal{O}(N^{-1/2}) \big)
\nn\\\simeq &
 \frac{\omega \sqrt{N}}{e^{2\pi \omega}-1}
\big(1+\mathcal{O}(N^{-1/2}) \big)
\,.
\end{align}
Hereafter, the $\mathcal{O}(N^{-1/2})$ correction terms,
$\mathcal{O}(1)$ numerical coefficients,
and the summation over the polarizations will often be implicit.
This leads a thermal distribution of the emission rate
\begin{align}
  \label{eq:12}
  \Gamma_B
\simeq &
\frac{\omega^8 d\omega}{M^2}
 \frac{\sqrt{N}}{e^{\beta_H \omega}-1}
\end{align}
with the inverse temperature $\beta_H=2\pi$, namely, the inverse Hagedorn temperature.

We move on to the fermion emission rate.
We have
\begin{align}
  &
P(\Phi_{N} \to u(k)+\Phi_{N'})
\nn\\=&
\frac{1}{4 \cG(N)}\sum_{u}
\oint \frac{dw}{2\pi i w}w^{-N}
\oint \frac{dv}{2\pi i v}v^{N-N'}
\left(
\sqrt{N} u^{a*} u^a
+\frac{u^{\da *} u^\da}{\sqrt{N}} \Omega(v,w)
\right) \Xi(v,w)
Z(w)
\label{eq:79}
\,.
\end{align}
There are two terms
in the parenthesis,
and it seems that the first term is dominant since we take
$N$ to be large.
It is indeed the case, as explicitly checked by evaluating
the contour integrals.
A brief comment on this comparison is found in the last part of
Appendix \ref{sec:eval-domin-contr}.
We, thus, focus on the first term.
In the same way as the boson case, we have
\begin{align}
P(\Phi_{N} \to u(k)+\Phi_{N'})
=&
\frac{\sqrt{N}}{4\cG(N)}\sum_{u} |u^a|^2 
\oint \frac{dw}{2\pi i w}
\frac{w^{-N'}}{1+w^{N-N'}}
Z(w)
\nn\\\simeq &
\frac{\sqrt{N}}{\cG(N)}\sum_{u} |u^a|^2 
\frac{e^{\pi\sqrt{8N'}} (N')^{-\frac{11}{4}}}
{1+e^{-\sqrt{2}\pi\frac{N-N'}{\sqrt{N'}}}}
\nn\\\simeq &
\sum_{u} |u^a|^2 
\frac{\sqrt{N}}
{e^{2\pi\omega}+1}
\,,
\end{align}
where the saddle point appears at the same value as the boson case,
$\beta=\pi\sqrt{2/N'}$.
Thus, we have the emission rate for massless fermion,
\begin{align}
  \Gamma_F
\simeq &
\frac{\omega^7 d\omega}{M^2}
 \frac{\sqrt{N}}{e^{\beta_H \omega}+1}
\,,
\end{align}
which depends on the same inverse temperature $\beta_H$.

We make a comment on the twisted trace part.
There is also contribution from nonplanar diagrams,
where the copies of the vertex operators are located on the opposite
ends of the open string world sheet.
The twisting is realized by the operator\cite{GSW}
\begin{align}
  \label{eq:25}
  \Theta = -(-1)^{\hat{N}}
\,,
\end{align}
for which the action on the vertex operator is
\begin{align}
  \label{eq:33}
  \Theta V'(k,z) \Theta = V'(k,-z)
\,,
\end{align}
where $V'(k,z)$ denotes the oscillator part of the vertex operator.
We need to include the twisted sector as in \cite{Amati:1999fv},
by replacing the vertex operator as $V(\gamma,k) \rightarrow
(V(\gamma,k) + \Theta V(\gamma,k) \Theta)/\sqrt{2}$.
We then have the untwisted part (taking the first vertex operator squared
or the second one squared), which is equivalent to the one we have already considered.
The other is the twisted part, which comes from the cross terms.
The net effect for the twisted part
is just to replace the location of the second
vertex operator as $V(\gamma,k,-v)$.
It amounts to replacing $\Omega(v,w) \rightarrow \Omega(-v,w)$ and
$\Xi(v,w) \rightarrow \Xi(-v,w)$ in evaluation of $v$ integral
in both the boson and the fermion emission rates.
Therefore, the final form is obtained by multiplying 
a level-difference-dependent phase factor to the
untwisted result as
\begin{align}
  \label{eq:12}
  \Gamma_B
\simeq &
\frac{\omega^8 d\omega}{M^2}
 \frac{(-1)^{N-N'} \sqrt{N}}{e^{\beta_H \omega}-1}
\,.
\end{align}
This tells that for odd $N - N'$, the twisted part contrition
will cancel out with the untwisted  one.
However, it does not change the thermal behavior of the decay rate,
and we simply omit the contribution from the twisted part.

In order to have a consistent
open-closed superstring theory in the flat spacetime of critical dimensions,
it is known that we need to consider
unoriented theory with an appropriate Chan-Paton factor.
For simplicity, we first examine the effect of unoriented projection 
without Chan-Paton factor.
The physical states are to satisfy the condition
\begin{align}
  \label{eq:42}
  \ket{\Phi} =& \frac{1+\Theta}{2} \ket{\Phi} \,.
\end{align}
By replacing the initial and the final physical state with these unoriented ones,
the calculations are parallel with the above twisted sector calculations.
Finally, we find, for example, for open bosonic emissions,
\begin{align}
  \Gamma_B
\simeq &
\frac{1+(-1)^{N-N'}-(-1)^{N}-(-1)^{N'}}{4}
\frac{\omega^8 d\omega}{M^2}
 \frac{\sqrt{N}}{e^{\beta_H \omega}-1}
\,.
\label{eq:Unori}
\end{align}
If both the initial-state level $N$ and the final one $N'$ are odd,
as required by the unoriented projection condition \eqref{eq:42},
the level-dependent phase factor of \eqref{eq:Unori}
 is trivially unity.
Therefore, it does not have any quantitative effect.
After including the Chan-Paton factor,
there appear both odd and even states.
They do not mix, and the decay rates for each of them are proportional to
oriented ones.
So it would make no difference as long as we are interested in
the thermal behavior
and we do not refer to the Chan-Paton factor and unoriented projection in this paper.

\subsection{Closed-string emission}
\label{sec:closed-string-case}

We move on to the consideration of emission of massless closed-string
states, namely, graviton, gravitino, dilaton, and so on.

\subsubsection{Closed string from a closed superstring}
\label{sec:closed-from-closed}

We consider the emission of a massless closed string state from a heavy
closed superstring.
The semi-inclusive decay process is the same as the heavy open string case.
The mass-shell condition for the closed string is
\begin{align}
  M^2 =& \frac{2}{\alpha'} \left(
N_R + N_L
\right)
= \frac{4}{\alpha'} N \,,
\end{align}
where $L$ and $R$ refer to the left- and right-moving part as usual.
From this, the level difference between the initial and the final
states is found to be
\begin{align}
  N-N' =& \sqrt{\alpha' N} \omega + \frac{\alpha'}{4}\omega^2 \,.
\end{align}
In this case, the momentum operator picks up the initial-state energy
$P_\text{ini}^+ = \sqrt{\frac{2N}{\alpha'}}$.

The calculation is parallel with the open-string case.
The oscillator part of the vertex operator is factorized as
\begin{align}
  \label{eq:34}
  V^\text{(closed)}(\gamma,k,e^{i\tau}) =&
\int_0^{\pi} \frac{d\sigma }{\pi}
:V_L(\gamma_L,\tfrac{k}{2},e^{i(\tau+\sigma)}): \, : V_R(\gamma_R,\tfrac{k}{2},e^{i(\tau-\sigma)}):
\nn\\=&
\int_0^{\pi} \frac{d\sigma }{\pi}
e^{-2i\sigma(\hat{N}_L-\hat{N}_R)}
:V_L(\gamma_L,\tfrac{k}{2},e^{i\tau}): \, : V_R(\gamma_R,\tfrac{k}{2},e^{i\tau}) :
e^{2i\sigma(\hat{N}_L-\hat{N}_R)}
\,,
\end{align}
where $V_{L,R}$ is either $V_B$ or $V_F$,
and $\gamma=\gamma_L \otimes \gamma_R$.
As shown, the normal ordering is taken for the left and right parts individually.
Since we consider massless vertex operators, we do not write the normal ordering
symbol hereafter.
As we consider a tree level three-point amplitude with closed string states
that satisfy the level matching condition,
the $\sigma$ integral trivially drops out.
The initial and final states are also decomposed into
\begin{align}
  \label{eq:43}
  \ket{\Phi(N)} = \ket{\Phi_L(N)} \otimes
\ket{\Phi_R(N)} \,,
\end{align}
and these two sectors must have the same level.
Then, the projection operator is also decomposed as
\begin{align}
  \label{eq:46}
  \hat{P}_N=& \oint \frac{dv_L}{2\pi i v_L} v_L^{\hat{N}_L-N}
\times \oint \frac{dv_R}{2\pi i v_R} v_R^{\hat{N}_R-N}\,,
\end{align}
which gives
\begin{align}
  \label{eq:47}
  \sum_{\Phi_L|N} \sum_{\Phi_R|N} \ket{\Phi_L(N)} \otimes
\ket{\Phi_R(N)}
=& \sum_{\Phi_L} \sum_{\Phi_R} 
\hat{P}_N \ket{\Phi_L} \otimes \ket{\Phi_R}
\end{align}
where in the sums on the right-hand side,
the levels of $\Phi_L$ and $\Phi_R$ are not restricted
to be the same.
As shown in Appendix \ref{sec:dens-stat-haged},
the density of states for closed string $\mathcal{G}^\text{cl}(N)$
is the
square of the open-string one.
Therefore, the probability can be evaluated as
\begin{align}
&  P(\Phi_N \rightarrow \gamma(k) + \Phi_{N'})
\nn\\=&
 \frac{1}{\mathcal{G}^\text{cl}(N)}
\sum_{\Phi |N} \sum_{\Phi |N'}
\sum_{\gamma_L, \gamma_R} \left|
\bra{\Phi(N')} V(\gamma,k, 1)
\ket{\Phi(N)}
\right|^2
\nn\\=&
 \frac{1}{\mathcal{G}^\text{cl}(N)}
\sum_{\Phi } \sum_{\Phi'}
\sum_{\gamma_L, \gamma_R} \left|
\bra{\Phi'}\hat{P}_{N'}  V(\gamma,k, 1)
\hat{P}_N \ket{\Phi}
\right|^2
\nn\\=&
 \frac{1}{\mathcal{G}(N)}
\int \frac{dv_L}{2\pi i v_L} v_L^{N-N'}
\int \frac{dw_L}{2\pi i w_L} w_L^{-N}
\sum_{\gamma_L} \tr \left(
V_L(\gamma_L,\tfrac{k}{2}, 1)^\dagger
V_L(\gamma_L,\tfrac{k}{2},  v_L) w_L^{\hat{N}_L}
\right)
\nn\\& \times
 \frac{1}{\mathcal{G}(N)}
\int \frac{dv_R}{2\pi i v_R} v_R^{N-N'}
\int \frac{dw_R}{2\pi i w_R} w_R^{-N}
\sum_{\gamma_R} \tr \left(
V_R(\gamma_R,\tfrac{k}{2},  1)^\dagger
V_R(\gamma_R,\tfrac{k}{2},  v_R) w_R^{\hat{N}_R}
\right)
\,.
\end{align}
After inserting the level projection operator, the calculation is
factorized
into the left- and right-moving parts.
By setting $\alpha'=2$, it is easy to see that each part is just a
copy of the amplitude of open-string one with $\alpha'=1/2$.
We define the contribution of the averaged trace
of bosonic and fermionic vertex operators
(with numerical factors neglected),
\begin{align}
  \label{eq:35}
  f_B =\sum_\zeta |\zeta^i|^2 \frac{\sqrt{N} \omega }{e^{2\pi \omega}-1}
\,,
\qquad
  f_F =\sum_u |u^a|^2 \frac{ \sqrt{N} }{e^{2\pi \omega}+1}
\,,
\end{align}
and the emission rate is
\begin{align}
  \label{eq:36}
  \Gamma^\text{cl}_{LR} = & \frac{\omega^7 d\omega}{M^2} f_L f_R
\,,
\end{align}
with $L,R$ being $B$ or $F$.

First, we consider the case
with $L=R=B$; namely,
we prepare the vertex operator for $\mathbf{8}_v \times \mathbf{8}_v$ states,
which include the graviton, dilaton, and $B$-field,
\begin{align}
  \label{eq:37}
  \Gamma^{\text{cl}}_{BB} \simeq &
 \frac{\omega^7 d\omega}{M^2} f_B f_B
=
 \sum_{\zeta^{ij}} (\zeta^{ij} \zeta^{ij*})
 \frac{\omega^8 d\omega}{M^2}
\frac{N\omega}{(e^{2\pi \omega}-1)^2}
\,.
\end{align}
In the same manner, $\mathbf{8}_c \times \mathbf{8}_s$ for
gravitino and dilatino and $\mathbf{8}_c \times \mathbf{8}_c$
for Ramond-Ramond (R-R) fields are given by
\begin{align}
  \label{eq:38}
  \Gamma^{\text{cl}}_{FB} \simeq &
 \frac{\omega^7 d\omega}{M^2} f_F f_B
=\sum_{u^{ia}} |u^{ia}|^2
 \frac{\omega^8 d\omega}{M^2}
\frac{N}{(e^{2\pi \omega}-1)(e^{2\pi \omega}+1)}
\,,
\end{align}
and
\begin{align}
  \label{eq:45}
  \Gamma^{\text{cl}}_{FF} \simeq 
 \frac{\omega^7 d\omega}{M^2} f_F f_F
=\sum_{\zeta^{ab}} |\zeta^{ab}|^2
 \frac{\omega^8 d\omega}{M^2}
\frac{N\omega^{-1}}{(e^{2\pi \omega}+1)^2}
\,,
\end{align}
respectively. 
For the type IIA closed-string case, the second fermionic
state is replaced with $\mathbf{8}_s$, 
but the result is essentially the same.
It should be noted that the thermal factors for the left and right mover,
\begin{align}
  \label{eq:39}
  \beta_L=\beta_R=2\pi = \pi \sqrt{2\alpha'} \,,
\end{align}
are half of the inverse Hagedorn temperature for the closed string,
\begin{align}
  \label{eq:40}
  \beta_H = \pi \sqrt{8\alpha'} = \beta_L + \beta_R \,,
\end{align}
since we are working with $\alpha'=2$.

\subsubsection{Closed-string emission  from an open superstring}
\label{sec:closed-from-open}

We consider a closed-string state emission from open-string states
and use the same closed-string vertex operator,
\begin{align}
  V^\text{(closed)}(\gamma,k,e^{i\tau}) =&
\int_0^{\pi} \frac{d\sigma }{\pi}
:V_L(\gamma_L,\tfrac{k}{2},e^{i(\tau+\sigma)}): \,
:  V_R(\gamma_R,\tfrac{k}{2},e^{i(\tau-\sigma)}):
\,,
\end{align}
but now both $V_L$ and $V_R$ include the same open-string oscillator
$\alpha_n^i$ and $S_n^a$.
We work with $\alpha'=1/2$ and
denote the position of the operator by $e^{i\sigma}$ (we take $\tau=0$).
By using the same trick,
we have
\begin{align}
  \label{eq:44}
&
P(\Phi_N \rightarrow\gamma(k)+\Phi_{N'})
\nn\\=&
\frac{1}{\mathcal{G}(N)}  \sum_{\Phi|N}    \sum_{\Phi|N'} \sum_{\gamma}
\big|\bra{\Phi(N')} V^\text{(closed)}(\gamma,k,1) \ket{\Phi(N)} \big|^2 
\nn\\=&
\frac{1}{\mathcal{G}(N)}
\int_0^\pi \frac{d\sigma}{\pi} \int_0^\pi \frac{d\sigma'}{\pi}
\oint \frac{dv}{2\pi v }v^{N-N'}  \oint \frac{dw}{2\pi w }w^{-N} 
\nn\\&\hskip2em \times
\tr \left[
V_R^\dagger(\gamma_R, \tfrac{k}{2},e^{-i\sigma'}) V_L^\dagger(\gamma_L, \tfrac{k}{2},e^{i\sigma'})
V_L(\gamma_L, \tfrac{k}{2},ve^{i\sigma}) V_R(\gamma_R, \tfrac{k}{2},ve^{-i\sigma})
w^{\hat{N}}\right]
\,.
\end{align}
Here $V_{L,R}$ is either $V_B$ or $V_F$.

Although we have now four vertex operators
inside the trace, the calculation is similarly
straightforward but lengthy.
The evaluation has been done in Appendix \ref{sec:closed-string-vertex},
and we cite the result in the following.
When $V_L=V_R=V_B$, namely, an NS--NS massless state emission case,
we find
\begin{align}
  \label{eq:48}
  \Gamma_{BB} \simeq
 \sum_{i,j} 
 \frac{\omega^8 d\omega}{M^2}
\frac{N\omega}{(e^{\pi \omega}-1)^2}
\times
\begin{cases}
  \zeta^{ij}(\zeta^{ij}+\zeta^{ji})^*
& N-N'= \text{even}
\\
  \zeta^{ij}(\zeta^{ij}-\zeta^{ji})^*
& N-N'= \text{odd}
\end{cases}
\,.
\end{align}
In this case, only the symmetric (antisymmetric) part is emitted
when the level difference is even (odd).
Note that the level difference is restricted to be even when we consider
the unoriented theory, and then this selection rule is consistent with
the unoriented projection.
For $V_L=V_F$ and $V_R=V_B$, namely, a massless fermionic state emission,
\begin{align}
  \label{eq:49}
  \Gamma_{FB}\simeq
 \sum_{i,a}
|\zeta^i|^2 |u^a|^2 
 \frac{\omega^8 d\omega}{M^2}
\frac{N}{(e^{\pi \omega}-1)(e^{\pi\omega}+1)}
\,,
\end{align}
where the numerical coefficients may be different for $N-N'$ even
or odd.
We are only interested in the $\omega$-dependent part of the emission rate
and $N$ dependence.
Finally, when $V_L=V_R=V_F$, an R--R boson emission rate is
\begin{align}
  \label{eq:50}
  \Gamma_{FF} \simeq
 \sum_{a,b}
 \frac{\omega^8 d\omega}{M^2}
\frac{\omega^{-1} N}{(e^{\pi\omega}+1)^2}
\times
\begin{cases}
  u^{ab}(u^{ab}-u^{ba})^*
& N-N'= \text{even}
\\
  u^{ab}(u^{ab}+u^{ba})^*
& N-N'= \text{odd}  
\end{cases}
\,.
\end{align}
Now the antisymmetric part is emitted when the level difference is even.
For fermionic sectors, the unoriented projection picks up the graded-symmetrized
states\cite{GSW}, and then it is again consistent with the unoriented projection.

Recall that we are working with $\alpha'=1/2$ here.
On the other hand, for closed-string emission from a heavy closed superstring,
we used $\alpha'=2$.
By taking this difference into account, one can find that
the $\omega$-dependent part of the emission rate is the same for heavy
open and closed superstrings.

\subsection{Summary of the results and discussion}
\label{sec:emiss-rate-massl}

The emission rate we have calculated so far can be written as
\begin{align}
  \Gamma
\simeq 
\frac{\omega^8 d\omega}{M^2}
 \frac{\sigma(\omega)}{e^{\beta_H \omega}\mp 1}
\,,  
\end{align}
where the $-$ sign is for massless boson emissions and the $+$ for fermionic ones.
$\beta_H=\pi\sqrt{8\alpha'}$ is the inverse Hagedorn temperature.
$\sigma(\omega)$ is the greybody factor and
$\sigma(\omega)=1$ means the pure blackbody radiation.
The results are at the leading order in the coupling constant $g_s$ and $1/N$
and are valid for $\sqrt{\alpha'} \omega \ll \sqrt{N}$.
We omit $\mathcal{O}(1)$  numerical coefficients,
and the summation over the polarizations is implicit.

All the information is now packed in $\sigma(\omega)$.
For massless open-string state emission, we have found
\begin{align}
  \label{eq:51}
  \sigma_B^{(\text{op})} =& g_s^2 \sqrt{N} \cdot 1 \,,
\qquad
  \sigma_F^{(\text{op})} = g_s^2 \sqrt{N} \cdot \omega^{-1} \,,
\end{align}
where in order to show explicitly that this is the leading order in the coupling constant,
we inserted the open-string coupling constant $g_s$
(we consider the amplitude squared).
$B$ and $F$ stand for boson and fermion massless state emission, respectively.
For massless closed string state emissions, both from heavy open and closed superstrings,
we have found
\begin{align}
  \label{eq:52}
  \sigma^{(\text{cl})}_{BB}=& g_s^4 N \cdot \frac{\omega (e^{\beta_H \omega}-1)}{(e^{\frac{\beta_H \omega}{2}}-1)^2}
\,,\\
  \sigma^{(\text{cl})}_{FB}=& g_s^4 N \cdot \frac {e^{\beta_H \omega}+1}{(e^{\frac{\beta_H \omega}{2}}-1)(e^{\frac{\beta_H \omega}{2}}+1)}
\,,\\
  \sigma^{(\text{cl})}_{FF}=& g_s^4 N \cdot \frac{\omega^{-1} (e^{\beta_H \omega}-1)}{(e^{\frac{\beta_H \omega}{2}}+1)^2}
\,.
\end{align}
Here, $BB$ stands for the massless states corresponding to $V_B \otimes V_B$ vertex operator,
and so on.

First of all, one should notice that
the greybody factors of massless closed string emissions
have the same form
regardless of whether its source is a heavy open superstring or a closed one.
This may be explained by the fact that to the leading order,
the emission of massless states is local on the world sheet.
So the emitted massless states are only affected by the excited level of
the heavy string but not by its topology.
Since open and closed superstrings have very different sets of states at a given level $N$,
it is interesting to see that the averaged states exhibit the same thermal behavior.
It should also be noted that $\sigma_{BB}$ has the same form as the greybody factor of
the near BPS D$1$--D$5$ black hole system\cite{Das:1996wn, Maldacena:1996ix},
with different $\beta$.
(This fact has already been pointed out by Amati and Russo\cite{Amati:1999fv} for the bosonic string case.)
As for fermion emissions, $\sigma_{FB}$ is slightly different from the one
calculated in \cite{Hosomichi:1997if} by a factor of $\omega$, which may be due to a
formal $s$-wave limit discussed later,
but the exponential factors are the same.
It should be interesting to study more on why this universal form appears.

The massless boson states of an open superstring show the blackbody spectrum.
The fermionic states have nontrivial $\omega^{-1}$ dependence, which might be interpreted as
an $s$-wave extrapolation of blackbody result shown below.
Intuitively, we may understand why an open string has a blackbody spectrum
in the following way.
The greybody factor
is identified with the absorption cross section.
Now, we  cast a massless open-string state from an asymptotic infinity
toward a heavy open superstring.
When the massless state is absorbed into the heavy string,
we observe the probability that the same state is reflected back from the string
with the same energy.
To the leading order, an open-string state can be captured or emitted only from the ends
of the heavy open string.
An open string may split at any point into two open strings,
but in order to emit a massless state, the interaction has to take place exactly at
each of the two ends.
The splitting probability of an open string is uniform\cite{splitting_prob},
 and then for a heavy and long open string
at level $N$, the probability of emitting massless states is suppressed by $1/N$ 
(or $2/N$ to be more precise).
Thus, for asymptotic observers, a heavy open string can be viewed as a hole of a cavity;
namely, once it absorbs a wave of a certain frequency, it will hardly reemit it.
We can then interpret the heavy open-string emission rate as a cavity radiation.
On the other hand, closed strings can emit massless states from any point of the world sheet.
So the probability is not dumped as the level gets higher, and it may have
a nontrivial greybody factor to the leading order in $1/N$.

It is also interesting to compare our result with
the greybody factors of black holes in higher dimensions.
In four dimensions, the greybody factors of spherically symmetric
black holes for bosons and fermions are calculated in \cite{Page:1976df,Unruh:1976fm}.
In higher dimensions, the formulas are derived by \cite{Harmark:2007jy, Kanti:2002nr, Kanti:2002ge}.
It can be schematically written as
\begin{align}
  \label{eq:29}
  \Gamma \simeq & \frac{\sigma_{j s}(\omega) \omega^8 
    d\omega}{e^{\beta\omega} \mp 1} \,,
\end{align}
in ten dimensions,
where $\sigma_{j s} (\omega)$ is the greybody factor
for spin-$s$ field.
$j$ denotes the total angular momentum of the partial wave,
and some examples for lower $s$ are
\begin{align}
  \label{eq:81}
  \sigma_{j0} = \omega^{2j} \,,
\qquad
\sigma_{j \frac{1}{2}} = \omega^{2j-1}
\,, \qquad
\sigma_{j 1} = \omega^{2j} \,.
\end{align}
Here, we write down only $\omega$ dependence
and neglect other factors including the dependence of the profile of the black hole.
It satisfies the constraint $j \geq s$, and
for the $\omega$ small region, the dominant contribution
comes from the modes $j=s$.
For the first few modes, the results are
\begin{align}
  \label{eq:53}
  \sigma_{00} = 1 \,,
\qquad
\sigma_{\frac{1}{2}\; \frac{1}{2}} = 1 \,,
\qquad
\sigma_{11} = \omega^2 \,,
\end{align}
and so on.
Note that in our calculation of open-string state emission,
the boson is a vector field, and the fermion is a Dirac field.
So the results do not agree with the black hole ones.
However, in our calculation, we integrate over the angular dependence of the decay rate,
and it essentially picks up the $s$-wave part ($j=0$ part) of the partial wave decomposition.
In the above formulas for black holes, if we take a formal limit of $j=0$, we get
\begin{align}
  \label{eq:54}
  \sigma_{\text{boson},\, j=0} = 1 \,,
\qquad
\sigma_{\text{fermion},\, j=0} = \omega^{-1} \,,
\end{align}
which depends only on whether $s$ is an integer or a half-integer.
This resembles our result for massless open-string state emission, but
we do not claim that this procedure is completely justifiable.
We leave further study on this suggestive observation to the future.

If we look at a particularly low-energy emission,
$\omega \ll T_H$, we can expand the
exponential factor to find that all the emission rates take the form of
\begin{align}
  \Gamma \sim \left( g_s^2 \sqrt{N}\right)^\alpha
 \frac{\omega^7 d\omega}{M^2}
\,,
\end{align}
with $\alpha=1$ for emission from a heavy open string
and $\alpha=2$ for that from a closed one.
In this regime, the law of equal partition works well, and 
there will be no distinction between bosonic and fermionic
emission.
We, thus, have
a thermodynamically acceptable result.

In the emission rates we have calculated, 
the coupling constant and the excited level $N$ of the heavy string
appears in the combination of $g_s^2 \sqrt{N}$.
As a $1/N$ expansion, the subleading corrections are
found to be $\mathcal{O}(N^{-1/2})$.
It would be interesting to investigate 
how the higher-order corrections in $g_s$ depend on $N$.
We may imagine that the subleading corrections appear in the same combination,
and if it were the case, the perturbative calculation is valid for
$g_s \ll N^{-1/4}$, which is smaller than the
corresponding point value $g_s \sim N^{-1/4}$ in the large-$N$ limit.
Namely, when $g_s^2 \sqrt{N} \simeq 1$, near the corresponding point,
the perturbative expansion of this type becomes invalid.
Another possibility is that the subleading corrections
have the same order in $N$, and the perturbative corrections
are negligible when $g_s \ll 1$.
In this case, when $g_s \sim N^{-1/2}$, the corrections become the same
order as the $1/N$ corrections of the leading order,
and a perturbative calculation might be useful even near the corresponding point. 
Of course, correction terms may appear in a completely different way.
However, the form of the higher-order corrections may tell us in what
regime of $g_s$ and $N$ 
we can use perturbation theory
and whether we may approach the black hole/string corresponding point.

\section{Conclusion}
\label{sec:conlusion-discussion}

In this paper, we have calculated the semi-inclusive decay rates of
very massive open and closed superstrings in the flat background
by use of the Green-Schwarz superstring in the light-cone gauge.
We focus on the emission of massless open- and closed-string states.
The initial state is averaged over all the states at a fixed level
$N$, and the final state is summed over.
In this setup, we find that the emission rates for all the cases,
open- and closed-string massless states from a heavy open superstring 
and closed string massless states from a heavy closed superstring,
exhibit the thermal distribution of the Hagedorn temperature
with possible greybody factor corrections.
In the thermal distribution at the Hagedorn temperature,
the dominant emission channel for an asymptotic observer
is due to massless states, and our result provides
the leading order of the emission spectrum of a heavy superstring.

It is notable that they have the same leading-order string coupling $g_s$
and the initial excited level $N$ dependence,
contrary to the suggestion in the previous literature.
It is also interesting that the greybody factors for massless closed-string emission
take the same form in the cases of decay from both heavy open and closed superstrings.

For open-string massless states from a heavy open superstring,
the emission rate of bosonic states exhibits the blackbody behavior, while
the fermion emission part involves a greybody factor $\sigma = \omega^{-1}$.
These behaviors approximate the $s$-wave approximation of the black hole greybody
factors but do not really agree with the greybody factors in the physical regime.
This result may suggest that a heavy open superstring would hardly reflect back the
incoming massless states once absorbed 
and would exhibit the thermal spectrum of the cavity radiation.

As for closed massless states emission, the greybody factors do not depend on
whether it is from heavy open superstring or closed one.
This suggests that these two heavy string states, as thermal equilibrium states,
 have a common essential property.
The frequency-dependent part of the
greybody factors also takes very similar forms to
 those of D1--D5 near BPS black holes.
So such a heavy superstring would also share the essential property with
this BPS black holes.
It will be interesting to study the origin of this similarity.

There will be a lot of future directions,
on top of the ones we have proposed so far here and
in Section \ref{sec:emiss-rate-massl}.
It is interesting to consider the scattering process
with a heavy string state.
This analysis should clarify if the ``greybody factor'' found here
indeed has the interpretation of the absorption cross section.
Another issue may be to observe the detailed angular dependence
of the heavy string decay.
The partial waves of massless states exhibit different behavior
in the case of the black hole.
It is, thus, worth carrying out a partial wave analysis
to observe that the angular dependence also gives a
consistent result or there appear some differences.

\section*{Acknowledgment}
\label{sec:acknowledgement}

The authors thank H.~Itoyama, H.~Kawai, and K.~Murakami for valuable discussions.
The authors also thank 
Center for Theoretical Sciences, Taipei, Taiwan, R.O.C.,
and Research Group for Mathematical Physics, Osaka City University, 
for warm hospitality.
The work of S.~K. is supported by NSC99--2112--M--029--003--MY3
and NSC101--2811--M--029--001.
The work of T.~M. is supported in part by JSPS Grand-in-Aid for Young Scientists No. 22740190.

\appendix
\section{Miscellaneous Calculations}
\label{sec:misc-calc}

\subsection{Density of states and Hagedorn temperature}
\label{sec:dens-stat-haged}

We are to evaluate the density of states at a high level in
the Green-Schwartz open superstring theory in the light-cone gauge.
The oscillators satisfy the standard (anti)commutation relations,
\begin{align}
  \label{eq:3}
  [\alpha_n^i , \alpha_m^j ]=& n \delta^{ij} \delta_{n+m}
\,,
\quad
\{S_n^a , S^b_m \}= \delta^{ab} \delta_{n+m}
\,,
\end{align}
and the level operator is $\hat{N} =\sum_{n=1}^\infty (\alpha_{-n}^i \alpha_{n}^i + n S_{-n}^a S_n^a)$.
We sometimes refer to the bosonic and the fermionic parts as $\hat{N}_B$ and $\hat{N}_F$
respectively.
We consider the following partition function:
\begin{align}
    Z(w)=&
\tr \, w^{\hat{N}} =
\sum_{n=1}^\infty {\cal G}(n)w^n
=
16 \left(\frac{{f}_+(w)}{f_-(w)} \right)^8
=16 \left[
\theta_4 \bigg( 0 \bigg| -\frac{i}{\pi}\ln w \bigg)
\right]^{-8}
\,,
\end{align}
where $16$ is the degeneracy of the ground state and
\begin{align}
  {f}_\pm (w) =\prod_{n=1}^\infty (1 \pm w^n) 
\end{align}
is the contribution of each fermionic ($+$) and bosonic ($-$) mode, respectively.
$\theta_4(\nu| \tau)$ is Jacobi's elliptic function,
\begin{align}
  \label{eq:55}
  \theta_4(\nu | \tau)=& \sum_{n=-\infty}^\infty (-1)^n q^{n^2} e^{2i n\pi  \nu}
\,,
\qquad (q=e^{i\pi \tau}) 
\end{align}
which enjoys the Modular transformation property\cite{GSW},
\begin{align}
  \label{eq:56}
  \theta_4(0|\tau)=& \left( -\frac{\ln q}{\pi} \right)^{-1/2} \theta_2(0|-1/\tau) \,,
\qquad
\theta_2(0|\tau)= 2 q^{1/4} f_-(q^2) \big(f_+(q^2) \big)^2 \,,
\end{align}
that is used below.

With $w=e^{-\beta}$, we calculate the density of states by
\begin{align}
    {\cal G}(N)=&
\oint \frac{dw}{2\pi i w} w^{-N} Z(w)
=\oint \frac{d\beta}{2\pi i} e^{N \beta}
Z(e^{-\beta})
\,,
\end{align}
and for large $N$, this can be evaluated by using the saddle-point method.
By use of the modular transformation property, one obtains 
\begin{align}
  \frac{f_-(w)}{{f}_+(w)}=&
 \left(
-\frac{\ln w}{\pi}
\right)^{-1/2}
2 \tilde{w}^{1/4}
f_-(\tilde{w}^2)
{f}_+(\tilde{w}^2)^2 ,
& (\tilde{w}=e^{-\pi/\beta})
\label{eq:27}
\,,
\end{align}
and then easily finds the saddle point,
\begin{align}
  \beta =& \pi \sqrt{\frac{2}{N}} \,.
\end{align}
After integrating out the Gaussian fluctuation,
we find
\begin{align}
  {\cal G}(N) \simeq &
2\sqrt[4]{2}
N^{-\frac{11}{4}}e^{\pi\sqrt{8N}} \,.
\end{align}
Using $M=\sqrt{N/\alpha'}$, this can be translated into
the asymptotic mass density,
\begin{align}
  \rho(M) \sim M^{-\frac{9}{2}} e^{\pi \sqrt{8\alpha'} M} \,,
\end{align}
which gives the inverse Hagedorn temperature $\beta_H= \pi
\sqrt{8\alpha'}$.

For closed string, the number operator is
$\hat{N}= \frac{1}{2} \left(
\hat{N}_L + \hat{N}_R \right)$,
and $\hat{N}_L$ and $\hat{N}_R$ are
for the left-moving oscillators $\alpha_n^i$ and $S_n^a$ and
the right ones $\tilde\alpha_n^i$ and $\tilde{S}_n^a$,
respectively.
Then, the closed-string level density appears to be the square of the open-string one,
\begin{align}
  {\cal G}^\text{cl}(N) = \left( {\cal G}(N) \right)^2
\,.
\end{align}
It is translated into the mass density as
\begin{align}
  \rho^\text{cl}(M) \sim M^{-10} e^{\pi \sqrt{8\alpha'}M}
\,.
\end{align}
The Hagedorn temperature of the closed superstring theory is, therefore, the same as
that of the open superstring theory.

\subsection{Evaluation of the probability}
\label{sec:calculation-trace}

\subsubsection{Open-string vertex operators in open superstring theory}
\label{sec:open-string-vertex}

We first present an explicit computation of the following traces:
\begin{align}
  \tr\left( V_B(\zeta,k,1)^\dagger V_B(\zeta,k,v) \, w^{\hat{N}}
  \right)
  \,,\qquad
  \tr\left( V_F(u,k,1)^\dagger V_F(u,k,v) \, w^{\hat{N}} \right)
\,,
\end{align}
where the vertex operators are\footnote{%
We follow the convention of \cite{GSW}.
The normalization of the fermionic oscillator is the one taken in Chapter 5
of it,
\begin{align}
  S^a(\tau) =& \frac{1}{\sqrt{2}} \sum_n S_{n}^a e^{-in\tau} \,.
\end{align}}
\begin{align}
   V _B(\zeta,k,z) =& \left(
\zeta^i(k) B^i - \zeta^-(k) {p}^+
\right) e^{i k \cdot X(z)} \,,
\\
  V_F(u,k,z) =& \left(
u^a(k) F^a + u^\da(k) F^\da
\right) e^{i k \cdot X(z)} \,,
\end{align}
with
\begin{align}
  B^i =& \dot{X}^i - R^{ij} k^j \,,
\\
F^a =& \sqrt{{p}^+} S^a \,,
\qquad
F^\da = \frac{1}{\sqrt{{p^+}}}
\left(
\gamma^i_{\da a} \dot{X}^i S^a 
+\frac{1}{3}: (\gamma^i S)^\da
R^{ij} : k^j
\right) ,
\end{align}
and
\begin{align}
  R^{ij}=& \frac{1}{2} S^a \gamma^{ij}_{ab} S^b
\end{align}
is the generator of the rotation.
In this subsection, we take $\alpha'=1/2$ and $p^+$ and $p^i$ are understood to be already
evaluated by the initial state values
$P_\text{ini}^+=\sqrt{N}$ and  $P_\text{ini}^i=0$ in the trace.
$\gamma^i_{\da a}$ and $\gamma^i_{a \da}$ are $8 \times 8$ matrices,
from which $SO(8)$ gamma matrices are constructed.
$\gamma^{ij}_{ab}$ is the usual antisymmetrized product of these matrices.
This form of the vertex operators is valid in the frame $k^+=0$.
As argued in the beginning of Section \ref{sec:open-string-emission},
our setup is consistent with this choice.
Because of this choice, the transverse polarization tensor $\zeta^i$
and 
chiral spinor $u^\da$, which have the sufficient degrees of freedom to represent
physical states in general, obey extra constraints,
\begin{align}
  k^i \zeta^i(k) =0 \,,
\qquad
\gamma^i_{a\da} k^i u^\da(k)=0 \,,
\end{align}
and then there is missing one (four) degree(s) of freedom for a boson
(fermion).
We then need to supply
\begin{align}
  \zeta^-=& \lim_{k^+\rightarrow 0} \frac{\zeta^i k^i}{k^+} \,,
\quad
u^a = \lim_{k^+\rightarrow0} -\frac{\gamma^i_{a\da} k^i u^\da(k)}{k^+} 
\end{align}
to fill out the degrees of freedom, and the vertex operators
include these degrees of freedom.

Let us calculate the $M$-point function of the following operator:
 \begin{align}
   \label{eq:16}
   V_\zeta(k,\rho)=&
 \exp \big( ik\cdot X(\rho) + \zeta \cdot \dot{X}(\rho) \big)
\end{align}
by use of the standard coherent state method\cite{GSW, Kuroki:2007aj}.
One finds
\begin{align}
&    \tr_B \left[
V_{\zeta_1} (k_1 , \rho_1) \cdots V_{\zeta_M}(k_M, \rho_M)
w^{\hat{N}_B}
\right]
\nn\\=&
f_-(w)^8
\exp\left[
\sum_{r<s} \left(
k_s\cdot k_r \ln \psi(c_{sr},w)
+(\zeta_s \cdot k_r - \zeta_r \cdot k_s) \eta(c_{sr},w)
+\zeta_s \cdot \zeta_r \Omega(c_{sr},w)
\right)
\right]
\label{eq:17}
\,,
\end{align}
where
\begin{align}
  \ln \psi(c,w) =&
-\sum_{n=1}^\infty \frac{c^n + (w/c)^n-2w^n}{n(1-w^n)} \,,
\\
\eta(c,w)=& -\sum_{n=1}^\infty \frac{c^n - (w/c)^n}{1-w^n} 
=c \frac{\partial}{\partial c} \ln \psi(c,w) \,,
\\
\Omega(c,w)=& \sum_{n=1}^\infty n \frac{c^n + (w/c)^n}{1-w^n} 
=-c \frac{\partial}{\partial c} \eta(c,w) \,.
\end{align}
They are the oscillator part of the corresponding functions for the standard
one-loop amplitudes\cite{GSW}.
The desired $M$-point function for $\zeta\cdot \dot{X} e^{ik\cdot X}$ 
is obtained by taking the linear term in $\zeta$.

In the calculation of the main part, the vertex operators are those
for massless states, $(k^i)^2= \zeta^i k^i=0$.
Together with these on-shell conditions,
we find
\begin{align}
  \label{eq:19}
  \tr_B \left(
e^{-ik\cdot X(1)} e^{ik\cdot X(v)} 
w^{\hat{N}_B}
\right)
=&
f_-(w)^{-8} \,, \\
  \tr_B \left(
e^{-ik\cdot X(1)} \dot{X}^i(v) e^{ik\cdot X(v)} 
w^{\hat{N}_B}
\right)
=&
  \tr_B \left(
\dot{X}^i(1) e^{-ik\cdot X(1)}  e^{ik\cdot X(v)} 
w^{\hat{N}_B}
\right) \nonumber \\
=&
-k^i \eta(v,w)
f_-(w)^{-8} \,, 
\\
  \tr_B \left(
\dot{X}^i(1) e^{-ik\cdot X(1)} \dot{X}^j(v) e^{ik\cdot X(v)} 
w^{\hat{N}_B}
\right)
=&
\delta^{ij} \Omega(v,w) 
f_-(w)^{-8} \,.
\end{align}

Next we move on to the fermionic oscillator part.
The basic trace is
\begin{align}
  \label{eq:20}
  \tr_F (w^{\hat{N}_F})=& 16 \,  {f}_+(w)^8 \,,
\end{align}
where $16$ comes from the vacuum degeneracy on the fermion zero modes.
We now insert the fermionic operators $S^a(z)$.
Since $(S_n^a )^2=0$ for $n \neq 0$, there needs to be the equal 
number of the raising oscillator $S_{-n}^a$ and the corresponding
lowering oscillator $S_n^a$ with $n>0$.
Therefore, for example,
\begin{align}
  \label{eq:21}
  \tr_F \left(
S^{a\dagger}(1) S^b(v) w^{\hat{N}_F}
\right)=&
\frac{1}{2} \sum_{n,m=-\infty}^\infty
  \tr_F \left(
S^{a}_{-n} S^b_m v^{-m}  \,  w^{\sum_{\ell} \ell
  S_{-\ell}^c S_\ell^c}
\right)
\nn\\=&
8\, \delta^{ab} {f}_+(w)^8 \Xi(v,w) \,,
\end{align}
where $\Xi(v,w)$ is defined in \eqref{eq:9}.
Because of $\delta^{ab}$, it is easy to see that
\begin{align}
  \label{eq:24}
  \tr_F (R^{ij} w^{\hat{N}_F} )=0 \,.
\end{align}
The other combinations of the fermionic operators that appear in our
trace
calculations are
 \begin{align}
 \tr_F \left[
S^a(1) : S^b R^{ij}:(v) w^{\hat{N}_F}
\right]
=&
 \gamma^{ij}_{cd} \,
\delta^{ad}\delta^{bc}
 f_+(w)^8
\Xi(v,w)
\,,\\
  \tr_F \left[
\left(  : S^b R^{ij}:(1) \right)^\dagger S^a(v) w^{\hat{N}_F}
\right]
=&
 \gamma^{ij}_{cd} \,
\delta^{ad}\delta^{bc}
 f_+(w)^8
\Xi(v,w)
\,,\\
     \tr_F \left[
\left(  : S^a R^{ij}:(1) \right)^\dagger
  : S^b R^{kl}:(v)
 w^{\hat{N}_F}
\right]
=&
\frac{1}{8} \gamma^{ij}_{cd}
\gamma^{kl}_{ef} 
 \delta^{df} \,
f_+(w)^8
\nn\\ & \times
\left(
\delta^{ac} \delta^{be}  \Xi(v,w) 
+ 2 (\delta^{ab}\delta^{ce} -2 \delta^{ae} \delta^{bc} ) \Xi(v,w)^3
\right)
\,.
\end{align}

With these preparations and by taking the conditions 
$k^ik^i = \zeta^i k^i=0$ into account,
the boson emission vertex trace is evaluated as
\begin{align}
  \label{eq:26}
  \tr\left( V_B(\zeta,k,1)^\dagger V_B(\zeta,k,v) \, w^{\hat{N}}
  \right)
=&  
 \left(
|\zeta^i|^2 \Omega(v,w)
+|\zeta^-|^2 ({p}^+)^2
\right)
Z(w)
\,,
\end{align}
where it is easy to see that the contributions involving
$R^{ij}k^j$ vanish.

For the fermion emission vertex part,
$\tr\left( V_F(u,k,1)^\dagger V_F(u,k,v) \, w^{\hat{N}} \right)$,
the part that is quadratic in $u^a(k)F^a$ is, for example,
\begin{align}
  \label{eq:28}
    \frac{p^+}{2} 
u^{a *} u^b
\tr_B \big(
e^{-i k \cdot X(1)} e^{i k \cdot X(v)}
w^{\hat{N}_B}
\big)
\tr_F\big(
S^a(1) S^b(v) w^{\hat{N}_F} 
\big)
=&
  \frac{1}{4}p^+
|u^a|^2
Z(w)
\Xi(v,w) \,.
\end{align}
The quadratic part of $F^\da$ takes a more complicated form, 
as seen in the above trace results,
but
it gets simplified by use of the equation of motion,
$\gamma^i_{a\da} k^i u^\da(k)=0$, 
and
\begin{align}
    \tr\left( V_F(u,k,1)^\dagger V_F(u,k,v) \, w^{\hat{N}} \right)
=&
\frac{1}{4} \bigg[  {p}^+
u^{a *} u^a
+ u^{\dot a *} \gamma_{\dot a b}^i u^{ b} {p}^i
+ u^{ a *} \gamma_{a\dot b}^i u^{\dot b} {p}^i
\nn\\&\hskip2em
+  \frac{ u^{\dot a *} u^{\dot a}}{ {p}^+}
\big(({p}^i)^2+  \Omega(v,w)  \big)
\bigg]
\Xi(v,w)
Z(w)
 \,.
\end{align}
The probabilities are evaluated in Section \ref{sec:open-string-emission}
by use of these trace expressions.

\subsubsection{Closed-string vertex operators in open superstring theory}
\label{sec:closed-string-vertex}

The closed-string vertex operator is given by
\begin{align}
  V^\text{(closed)}(\gamma,k,e^{i\tau}) =&
\int_0^{\pi} \frac{d\sigma }{\pi}
:V_L(\gamma_L,k_L,e^{i(\tau+\sigma)}): :V_R(\gamma_R,k_R,e^{i(\tau-\sigma)}):
\,,
\end{align}
where $k_L=k_R=\frac{k}{2}$
and $\gamma= \gamma_L \otimes \gamma_R$.
$V_L$ and $V_R$ are either $V_B$ or $V_F$ of the open superstring vertex operator,
each of which is normal ordered as shown, but the whole is not.
We have the following three kinds of probabilities.

\paragraph{$BB$ part:}
\label{sec:bb-part}

The vertex operator is
\begin{align}
  {\cal V}_{BB} (\zeta \otimes \bar\zeta, k , e^{i\tau})
=& \int_0^\pi \frac{d\sigma}{\pi}
V_B(\zeta, k_L, e^{i( \tau+\sigma)})
V_B(\bar\zeta, k_R, e^{i(\tau-\sigma)})
\,,
\end{align}
and the probability is given by
\begin{align}
&
 \frac{1}{\mathcal{G}(N)}
 \oint \frac{dv}{2\pi i v} v^{N-N'}
 \oint \frac{dw}{2\pi i w} w^{-N}
 \tr \left(
 {\cal V}_{BB}^\dagger (\zeta \otimes \bar\zeta, k,1)
 {\cal V}_{BB}(\zeta \otimes \bar\zeta,k, v)
 w^{\hat{N}}
 \right)
\nn\\=&
\frac{1}{\mathcal{G}(N)}  \int_0^\pi \frac{d\sigma}{\pi}
  \int_0^\pi \frac{d\rho}{\pi}
\oint \frac{dv}{2\pi i v} v^{N-N'}
\oint \frac{dw}{2\pi i w} w^{-N}
\nn\\& \times
\tr \bigg[
\big( (\bar\zeta^{*i} B^i - \bar\zeta^{*-} p^+) e^{-ik_R \cdot X}(c_1) \big)
\big( (\zeta^{*j} B^j - \zeta^{*-} p^+) e^{-ik_L \cdot X}(c_2) \big)
 \nn\\& \hskip4em \times
\big( (\zeta^k B^k - \zeta^- p^+) e^{ik_L \cdot X}(c_3) \big)
\big( (\bar\zeta^l B^l - \bar\zeta^- p^+) e^{ik_R \cdot X}(c_4) \big)
w^{\hat{N}} \bigg]
\,,
\label{eq:23}
\end{align}
where
$c_1= e^{-i\rho}$,
$c_2=e^{i\rho}$,
$c_3=ve^{i\sigma}$,
and
$c_4=ve^{-i\sigma}$.
Since $(k^i)^2 = k^i \zeta^i =k^i\bar\zeta^i=0$, one can easily see that the terms involving
$R^{ij} k^j$ inside $B^i$ do not contribute to the trace.
We temporarily neglect the terms including $\zeta^-$, which will turn out to be subleading
in $1/N$.
Therefore, this trace is essentially the $M=4$ case of \eqref{eq:17} with four $\dot{X}$ insertions.
Therefore,
\begin{align}
  \eqref{eq:23} =&
  \int_0^\pi \frac{d\sigma}{\pi}
  \int_0^\pi \frac{d\rho}{\pi}
\oint \frac{dv}{2\pi i v} v^{N-N'}
\oint \frac{dw}{2\pi i w} w^{-N}
\left(
|\zeta \cdot \bar\zeta|^2
\Omega(c_{21},w) \Omega(c_{43},w)
\right.\nn\\&\left.\hskip3em
+|\zeta \cdot \bar\zeta^*|^2
\Omega(c_{31},w) \Omega(c_{42},w)
+|\zeta|^2 |\bar\zeta|^2
\Omega(c_{41},w) \Omega(c_{32},w)
\right)
Z(w)
\,,
  \label{eq:18}
\end{align}
where $c_{rs} = c_r/ c_s$.
The $v$ integral again picks up the terms including $v^{-(N-N')}$ with $N-N'>0$.
So the first term drops out since $c_{21}$ and $c_{43}$ are independent of $v$.
We have a double sum from the quadratic of $\Omega(v,w)$,
and after the $v$ integration, we are left with a single sum over positive integer $n$.
$\rho$ and $\sigma$ integrals give a projection condition for $N-N'$, and we get
\begin{align}
& \eqref{eq:18}
\nn\\  =&
 \frac{1}{\mathcal{G}(N)}
\oint \frac{dw}{2\pi i w} w^{-N}
\left(
\zeta^{ij}(\zeta^{ij *}+\zeta^{ji *} )
P_1(w)
\frac{1+(-1)^{L}}{2}
+\zeta^{ij}(\zeta^{ij *}-\zeta^{ji *} )
\frac{2}{\pi^2}P_2(w)
\frac{1-(-1)^{L}}{2}
\right) 
Z(w)
\,,
  \label{eq:31}
\end{align}
where $L=N-N' = \sqrt{2 N} \omega + \mathcal{O}(1) >0$
and
\begin{align}
  P_1(w)=&
  \left( \frac{L}{2} \right)^2
\frac{w^{L}}{\left( 1-w^{\frac{L}{2}} \right)^2}
\,,\\
P_2(w)=&
2\sum_{n=1}^\infty 
\frac{n(n+L)}{(2n+L)^2} \frac{w^{L+n}}{(1-w^n)(1-w^{n+L})}
+\sum_{n=1}^{L-1 }
\frac{n(L-n)}{(2n-L)^2} \frac{w^{L}}{(1-w^n)(1-w^{L-n})}
\label{eq:69}
\,.
\end{align}
We have also defined $\zeta^{ij} = \zeta^i \otimes \bar \zeta^j$.
So far, we have not relied on any approximation.
We are going to evaluate the $w$ integral by the saddle-point method as before.
In the $P_2(w)$ part, we need to further evaluate where the dominant
contribution comes from.
As examined in Section \ref{sec:eval-domin-contr},
 it turns out that the dominant contribution comes 
 from the second finite sum with $n=(L+1)/2 +k$,
 where $k$ runs over an $\mathcal{O}(1)$ region,
and we concentrate on this contribution.
For the $P_1$ part in \eqref{eq:31}, the $w$ integral is very similar to the open-string
 vertex operator case
($L$ is assumed to be even),
\begin{align}
  \label{eq:32}
\zeta^{ij}(\zeta^{ij *}+\zeta^{ji *} )
 \frac{1}{\mathcal{G}(N)}
\oint \frac{dw}{2\pi i w} w^{-N}
P_1(w)
Z(w)
\simeq &
\zeta^{ij}(\zeta^{ij *}+\zeta^{ji *} )
\frac{N\omega^2 }{(e^{\pi\omega}-1)^2} \,.
\end{align}
For the $P_2$ part, we consider only the second sum with $n=(L+1)/2 +k$
and find ($L$ is assumed to be odd)
\begin{align}
  \label{eq:57}
  &
\sum_{k=-\mathcal{O}(1)}^{\mathcal{O}(1)}
2 \zeta^{ij} \zeta^{*[ij]} \frac{2}{\pi^2}
\frac{L^2/4}{(2k+1)^2}
  \frac{1}{\mathcal{G}(N)} \oint \frac{dw}{2\pi i w} w^{-N}
\frac{w^L}{(1-w^{\frac{L}{2}+k+\frac{1}{2}})(1-w^{\frac{L}{2}-k-\frac{1}{2}})}
Z(w)
\nn\\ \simeq &
\frac{1}{2\pi^2 }\sum_{k=-\mathcal{O}(1)}^{\mathcal{O}(1)}
2 \zeta^{ij} \zeta^{*[ij]}
\frac{(N-N')^2}{(2k+1)^2}
\frac{e^{-2\pi\omega}}{(1-e^{-\pi\omega-(k+\frac{1}{2})\pi\sqrt{\frac{2}{N'}}})
(1-e^{-\pi\omega+(k+\frac{1}{2})\pi\sqrt{\frac{2}{N'}}})}
\nn\\ \simeq &
2 \zeta^{ij} \zeta^{*[ij]}
\frac{N\omega^2}{(e^{\pi\omega}-1)^2}
\sum_{k=-\mathcal{O}(1)}^{\mathcal{O}(1)}
\frac{1/2\pi^2}{(2k+1)^2}
\nn\\ \simeq &
\zeta^{ij} \zeta^{*[ij]}
\frac{N\omega^2 }{(e^{\pi\omega}-1)^2}
\,,
\end{align}
where we have omitted an $\mathcal{O}(1)$ numerical coefficient.
Thus up to a numerical coefficient, the probability is the same as 
the $L$ even part.

Now we come back to the part that depends on $\zeta^-$.
Due to $(k^i)^2=\zeta^ik^i=\bar\zeta^i k^i=0$, it is easy to see that
only the terms with an even number of $\zeta^-$ survive in the trace.
The quartic term in $\zeta^-$ is independent of $v$, and then it drops
out after the $v$ integral.
The only relevant terms are quadratic in $\zeta^-$, and
\begin{align}
&
  \tr
\big(
\mathcal{V}_{BB}^\dagger(k,1) \mathcal{V}_{BB}(k,v) w^{\hat{N}} \big)
 \bigg|_{\text{quadratic in }\zeta^-} 
\nn\\=&
N \bar\zeta^{-*} \zeta^- \zeta^{i*} \bar\zeta^i \Omega(c_{42},w) 
+|\bar\zeta^{-}|^2 |\zeta^i|^2 \Omega(c_{32},w)
+|\zeta^-|^2 |\bar\zeta^i|^2 \Omega(c_{41},w)
+\zeta^{-*} \bar\zeta^- \bar\zeta^{i*} \zeta^i \Omega(c_{31},w)
\,.
  \label{eq:58}
\end{align}
The integrals with respect to $v$, $\rho$, $\sigma$ and $w$ are easily carried out,
 and it turns out that these terms are of order one, and, thus, are subleading contributions.

\paragraph{$FB$ part:}
\label{sec:fb-part}

The vertex operator is
\begin{align}
  \mathcal{V}_{FB}(u\otimes \zeta,k,e^{i\tau})=&
\int_0^\pi \frac{d\sigma}{\pi} \,
V_{F}(u,k_L; e^{i(\tau+\sigma)}) 
V_{B}(\zeta,k_R; e^{i(\tau-\sigma)}) 
\end{align}
and the probability is given by
\begin{align}
&
\frac{1}{\mathcal{G}(N)}
\oint \frac{dv}{2\pi i v} v^{N-N'}
\oint \frac{dw}{2\pi i w} w^{-N}
\tr \left(
{\cal V}_{FB}^\dagger (u\otimes \zeta, k, 1)
{\cal V}_{FB}(u\otimes \zeta, k, v )
w^{\hat{N}}
\right)
\nn\\=&
\frac{1}{\mathcal{G}(N)}  \int_0^\pi \frac{d\sigma}{\pi}
  \int_0^\pi \frac{d\rho}{\pi}
\oint \frac{dv}{2\pi i v} v^{N-N'}
\oint \frac{dw}{2\pi i w} w^{-N}
\nn\\&\hskip2em
\times \tr \bigg[
(\zeta^{i*} {B}^{i\dagger} + \zeta^{-*} p^+ )e^{- ik_R \cdot X}(c_1)
(u^{a*} F^{a\dagger} + u^{\da *} F^{\da \dagger} )e^{-ik_L \cdot X}(c_2)
\nn\\&\hskip3em\times
(u^b F^b + u^{\dot b} F^{\dot b} )e^{ik_L \cdot X}(c_3)
(\zeta^j {B}^j + \zeta^- p^+ )e^{ik_R \cdot X}(c_4)
w^{\hat{N}}
\bigg]
\,.
\end{align}
Here, we will neglect the contributions from
$F^{\da}$ parts which is $1/p^+ = N^{-1/2}$ smaller than $F^a$
and would be subleading.
For open-string massless states emission, this fact has been explicitly confirmed.
$c_1,\cdots,c_4$ are the same ones in the $BB$ part.
For the bosonic vertex operator part, the terms including an odd number of $\zeta^- p^+$ 
vanish due to the equation of motion of $u^a$.
The terms like $R^{ij} k^k$ can also be omitted due to $(k^i)^2= \zeta^i k^i=0$.
We then have two kinds of terms.
First, for terms of the quadratic in $\zeta^- p^+$, the trace is evaluated as
\begin{align}
  \label{eq:59}
  &
 |\zeta^- p^+|^2 u^{*a} u^b
 \tr_B \left(
e^{-\frac{i}{2}k\cdot X}(c_1)
e^{\frac{i}{2}k\cdot X}(c_4)
w^{\hat{N}_B} \right)
\cdot
p^+ 
\tr_F \left(
S^a(c_2) S^b (c_3)
w^{\hat{N}_F}
\right)
\nn\\=&
\frac{1}{4} |\zeta^-|^2 N^{3/2}
|u^a|^2
\Xi(c_{41},w) 
Z(w)
\end{align}
and the explicit evaluation of integrals shows that
it is $\mathcal{O}(N^{-1/2})$ and negligible.
The other term is
\begin{align}
  \label{eq:60}
&
\frac{1}{\mathcal{G}(N)}  \int_0^\pi \frac{d\sigma}{\pi}
  \int_0^\pi \frac{d\rho}{\pi}
\oint \frac{dv}{2\pi i v} v^{N-N'}
\oint \frac{dw}{2\pi i w} w^{-N}
\nn\\&\hskip2em \times
 \zeta^{*i}\zeta^j u^{*a} u^b
 \tr_B \left(
\dot{X}^i e^{-\frac{i}{2}k\cdot X}(c_1)
\dot{X}^j e^{\frac{i}{2}k\cdot X}(c_4)
w^{\hat{N}_B} \right)
\cdot
p^+ 
\tr_F \left(
S^a(c_2) S^b (c_3)
w^{\hat{N}_F}
\right)  
\nn\\=&
\frac{\sqrt{N} |\zeta^i|^2 |u^a|^2}{4 \mathcal{G}(N)}  \int_0^\pi \frac{d\sigma}{\pi}
  \int_0^\pi \frac{d\rho}{\pi}
\oint \frac{dv}{2\pi i v} v^{N-N'}
\oint \frac{dw}{2\pi i w} w^{-N}
  \Omega(c_{41},w) \Xi(c_{32},w)
Z(w)
\nn\\=&
\frac{\sqrt{N}}{4} |\zeta^i|^2 |u^a|^2
\oint \frac{dw}{2\pi i w} w^{-N}
\left[
\bar{P}_1(w) \frac{1+(-1)^L}{2}
+\frac{4}{\pi^2} \bar{P}_2(w) \frac{1-(-1)^L}{2}
\right] \, Z(w)
\,,
\end{align}
where $L=N-N'$ and
\begin{align}
  \label{eq:61}
  \bar{P}_1(w) =&
\frac{L}{2} \frac{w^L}{1-w^L} \,,
\\
\bar{P}_2(w) =&
2 \sum_{n=1}^\infty 
\frac{n}{(2n+L)^2} \frac{w^{L+n}}{(1-w^n)(1+w^{L+n})}
+ \sum_{n=1}^{L-1} 
\frac{n}{(2n-L)^2} \frac{w^{L}}{(1-w^n)(1+w^{L-n})}
+\frac{1}{2L} \frac{w^L}{1-w^L}
\label{eq:77}
\,.
\end{align}
For $\bar{P}_2(w)$ part, the dominant contribution
comes from again $n=(L+1)/2+k$ with $k=\mathcal{O}(1)$
of the second sum,
and after evaluating all the integrals, one gets
\begin{align}
\eqref{eq:60} \simeq &
|\zeta^i|^2 |u^a|^2 \frac{N\omega}{e^{2\pi\omega}-1} \,,
  \label{eq:62} 
\end{align}
where $L$-even and -odd parts would have different $\mathcal{O}(1)$ numerical coefficients.

\paragraph{$FF$ part:}
\label{sec:ff-part}

The vertex operator we use is
\begin{align}
  \mathcal{V}_{FF}(u\otimes \bar u, k, e^{i\tau})=&
\int_0^\pi \frac{d\sigma}{\pi} \,
V_{F}(u,k_L, e^{i(\tau+\sigma)}) 
V_{F}(\bar u,k_R, e^{i(\tau-\sigma)}) 
\end{align}
and we will evaluate
\begin{align}
&
\frac{1}{\mathcal{G}(N)} 
\oint \frac{dv}{2\pi i v} v^{N-N'}
\oint \frac{dw}{2\pi i w} w^{-N}
\tr \left(
{\cal V}_{FF}^\dagger (u\otimes \bar u, k,1 )
{\cal V}_{FF}(u\otimes \bar u, k, v )
w^{\hat{N}}
\right)
\nn\\=&
\frac{1}{\mathcal{G}(N)} 
  \int_0^\pi \frac{d\sigma}{\pi}
  \int_0^\pi \frac{d\rho}{\pi}
\oint \frac{dv}{2\pi i v} v^{N-N'}
\oint \frac{dw}{2\pi i w} w^{-N}
(p^+)^2 \bar{u}^{*a}(k_R) u^{*b}(k_L) u^{c}(k_L) \bar{u}^d(k_R)
\nn\\&\hskip2em \times
(f_-(w))^{-8} \, 
\tr_F \bigg[
S^a (c_1)
S^b (c_2)
S^c (c_3)
S^d (c_4)
w^{\hat{N}_F}
\bigg]
\,,
\label{eq:65}
\end{align}
where we have omitted again the subleading terms $F^\da$.
All we need to do is to evaluate this fermionic trace, and
after some algebra, we find
\begin{align}
  \label{eq:63}
  &    \tr_F \left(
S^a (c_1)
S^b (c_2)
S^c (c_3)
S^d (c_4)
w^{\hat{N}_F}
\right)
\nn\\=&
4(f_+(w))^8
\bigg[  
 \delta^{ab}\delta^{bc}\delta^{cd}
\bigg(
-{1\over2}+ \Xi(c_{21}c_{43},w)
-\sum_{m=1}^\infty
\frac{\left( c_{21}^m + (w/c_{21})^m \right) \left(c_{43}^m + (w/c_{43})^m \right)}{(1+w^m)^2}
\nn\\&
+\sum_{m=1}^\infty
\frac{\left( c_{31}^m + (w/c_{31})^m \right) \left(c_{42}^m + (w/c_{42})^m \right)}{(1+w^m)^2}
-\sum_{m=1}^\infty
\frac{\left( c_{41}^m + (w/c_{41})^m \right) \left(c_{32}^m + (w/c_{32})^m \right)}{(1+w^m)^2}
\bigg)
\nn\\&
+\delta^{ab}\delta^{cd}\Xi(c_{43},w)\Xi(c_{21},w) 
-\delta^{ac}\delta^{bd}
\Xi(c_{42},w)\Xi(c_{31},w)
+\delta^{ad}\delta^{bc}
\Xi(c_{41},w)\Xi(c_{32},w)
\bigg]
\,.
\end{align}
Here, $c_{21}$ and $c_{43}$ do not depend on $v$, and the terms only including these two variables
vanish after the $v$ integral.
The other $\delta^{ab} \delta^{bc} \delta^{cd}$ part turns out to cancel exactly after
the $v$, $\rho$, and $\sigma$ integrals.
Therefore, we finally have
\begin{align}
\eqref{eq:65} =&
\frac{N}{4} \bar{u}^{*a} u^{*b} u^{c} \bar{u}^d
\oint \frac{dw}{2\pi i w} w^{-N}
Z(w)
\nn\\& \hskip2em \times 
\bigg[
\left(\delta^{ad} \delta^{bc} - \delta^{ac} \delta^{bd} \right)
\tilde{P}_1(w) \frac{1+(-1)^{L}}{2}
+\left(\delta^{ad} \delta^{bc} + \delta^{ac} \delta^{bd} \right)
\frac{4}{\pi^2} 
\tilde{P}_2(w) \frac{1-(-1)^{L}}{2}
\bigg]
\,,
  \label{eq:64}  
\end{align}
where $L=N-N'$ and
\begin{align}
  \label{eq:66}
  \tilde{P}_1(w)=& \frac{w^{L}}{\left( 1+w^{\frac{L}{2}} \right)^2} \,,
\\
\tilde{P}_2(w)=&
2\sum_{n=1}^\infty 
\frac{1}{(2n+L)^2} \frac{w^{n+L}}{(1+w^n)(1+w^{L+n})}
+ \sum_{n=1}^{L-1 }
\frac{1}{(2n-L)^2} \frac{w^{L}}{(1+w^n)(1+w^{L-n})}
+\frac{1}{L^2} \frac{w^{L}}{1+w^{L}}
\label{eq:78}
 \,.
\end{align}
Again, only the $n=(L+1)/2+k$ part of the second sum
gives the dominant contribution in
$\tilde{P}_2(w)$.
The parts in which $L$ is even and odd are evaluated as
\begin{align}
\eqref{eq:64} =&
C  u^{ab}(u^{ab}-u^{ba}) \frac{N}{(e^{\pi\omega}+1)^2}
\frac{1+(-1)^{L}}{2}
+C'  u^{ab}(u^{ab}+u^{ba}) \frac{N}{(e^{\pi\omega}+1)^2}
\frac{1-(-1)^{L}}{2}
\,,
  \label{eq:67}
\end{align}
where $C,C'$ are certain $\mathcal{O}(1)$ numerical coefficients.
We have defined $u^{ab} = u^a \otimes \bar{u}^b$.

\subsubsection{Evaluation of the dominant contribution in the summation}
\label{sec:eval-domin-contr}

When we evaluate the probability $P(\Phi_N \rightarrow \gamma(k) + \Phi_{N'})$,
 we sometimes need to evaluate an infinite sum, such as
\eqref{eq:69} and to figure out from which part of the sum
the dominant contribution comes.
We now take the $P_2(w)$ that appears in the $BB$ part calculation as
an example and demonstrate that the $n=(L+1)/2+k$ part gives a leading contribution.
We evaluate the $w$ integral of $P_2(w)$ (defined by \eqref{eq:69}),
\begin{align}
&  \frac{1}{\mathcal{G}(N)} \oint \frac{dw}{2\pi i w} w^{-N}
\bigg[
2\sum_{n=1}^\infty 
\frac{n(n+L)}{(2n+L)^2} \frac{w^{L+n}}{(1-w^n)(1-w^{n+L})}
+\sum_{n=1}^{L-1 }
\frac{n(L-n)}{(2n-L)^2} \frac{w^{L}}{(1-w^n)(1-w^{L-n})}
\bigg]
Z(w)
\,.
  \label{eq:70}
\end{align}
Recall that $L=N-N'>0$ and only the $L$-odd case is relevant.
We set $w=e^{-\beta}$ and evaluate the $\beta$ integral by the saddle-point method.
First, we consider the $n \gg N$ case.
In this case, the saddle point is determined by the $w^n$ part,
which is now at $\beta=\pi\sqrt{2/n}$,
and after the saddle-point approximation,
\begin{align}
\eqref{eq:70} \simeq &
\frac{1}{\mathcal{G}(N)} n^{-\frac{11}{4}}  e^{\pi\sqrt{8n}}
\sum_{n \gg N}
\frac{e^{\frac{N'\pi\sqrt{2}}{\sqrt{n}}}}{(e^{{\sqrt{2n}\pi}}-1)
(e^{{\sqrt{2n}\pi}}-e^{\frac{L\pi\sqrt{2}}{n}})}  
\nn\\ \simeq &
\left(  \frac{N}{n} \right)^{\frac{11}{4}}
\exp \left(
\frac{N'\pi\sqrt{2}}{\sqrt{n}}
-\pi\sqrt{8N}
 \right)
\sum_{n \gg N}
\frac{1}{(1-e^{-{\sqrt{2n}\pi}})
(1-e^{\frac{(L-n)\pi\sqrt{2}}{n}})}  
\,.
  \label{eq:71} 
\end{align}
The overall factor is small like $e^{-\sqrt{N}}$, and then this part has an exponentially small
contribution for large $N$.

Next, we consider the case with $n=\mathcal{O}(N)$.
The saddle point is shifted due to the $w^n$ part to
$\beta=\pi \sqrt{2/(N'+n)}$.
After the saddle-point approximation, we find
\begin{align}
\eqref{eq:70} \simeq &
  \left( \frac{N'+n}{N} \right)^{-\frac{11}{4}}
 e^{\pi\sqrt{8(N'+n)}-\frac{2n\pi\sqrt{2}}{\sqrt{N'+n}}-\pi\sqrt{8N}}
\sum_{n\sim \mathcal{O}(N)}
\frac{1}{(1-e^{-\frac{n\pi\sqrt{2}}{\sqrt{N'+n}}})
(1-e^{\frac{(L-n)\pi\sqrt{2}}{\sqrt{N'+n}}})}
\,.
  \label{eq:72}
\end{align}
The factor inside the sum is not large.
We take $n=c N$ with $c=\mathcal{O}(1)>0$ and evaluate the overall exponential factor,
\begin{align}
  \label{eq:73}
    \exp \left(
\pi\sqrt{8(N'+n)}-\frac{n\pi\sqrt{8}}{\sqrt{N'+n}} - \pi\sqrt{8N}
\right)
=
\exp\left(
\pi\frac{1-\sqrt{1+c}}{\sqrt{1+c}}\sqrt{8N}
+\mathcal{O}(1)
\right)
\,.
\end{align}
Since $c>0$, we find that this factor is exponentially small for large $N$.
Therefore, we conclude that
the $n=\mathcal{O}(N)$ part does not give a leading-order contribution.

Finally, we evaluate the case with $n \ll N$.
In this case, the saddle point for the $\beta$ integral is the same as before,
$\beta=\pi \sqrt{2/N'}$.
After evaluating the integral, we have
\begin{align}
\eqref{eq:70} \simeq &
e^{-2\pi\omega}
\sum_{n \ll N} \left[
\frac{2n(n+L)}{(2n+L)^2} \frac{e^{-\pi\sqrt{2} \frac{n}{\sqrt{N'}}}}{(1-e^{-\pi\sqrt{2} \frac{n}{\sqrt{N'}}})(1-e^{-2\pi\omega-\pi\sqrt{2} \frac{n}{\sqrt{N'}}})}
\right.\nn\\&\left.\hskip6em
+
\frac{n(L-n)}{(2n-L)^2} \frac{1}{(1-e^{-\pi\sqrt{2} \frac{n}{\sqrt{N'}}})(1-e^{-2\pi\omega + \pi\sqrt{2} \frac{n}{\sqrt{N'}}})}
\right]
\,,
  \label{eq:74}
\end{align}
where the overall factor in front of the sum is $\mathcal{O}(1)$.
We evaluate the sum in order:
\begin{enumerate}
\item $n=\mathcal{O}(1)$:
The summation can be written as
\begin{align}
  \label{eq:75}
  \sum_{n=\mathcal{O}(1)}  \mathcal{O}(N^{-1/2}) \frac{\sqrt{N}/n}{1-e^{-2\pi\omega}}
\end{align}
where one of the exponential factor is expanded.
Each term is $\mathcal{O}(1)$ and there are in total $\mathcal{O}(1)$ number of 
terms.
So the contribution from this part will be $\mathcal{O}(1)$.
\item $\sqrt{N} \ll n \ll N$:
The second sum does not have this range of $n$.
The first sum has an exponential dumping factor $e^{-n/\sqrt{N}}$,
and then this part has an exponentially small contribution.
\item $n=\mathcal{O}(\sqrt{N})$:
Inside the sum, each of the exponential term is $\mathcal{O}(1)$.
The factor in front of it is, in general,
\begin{align}
  \label{eq:76}
  \frac{n(L \pm n)}{(2n \pm L)^2} \bigg|_{n=\mathcal{O}(\sqrt{N})}
= \mathcal{O}(1) 
\end{align}
since $L=\mathcal{O}(\sqrt{N})$.
The number of the sum is much smaller than $\mathcal{O}(N)$ and then the contribution
is much smaller than $\mathcal{O}(N)$.
However, in the second sum, when $n=(L+1)/2 +k$, the above factor gets enhanced 
and becomes $\mathcal{O}(N)$.
$k$ is $\mathcal{O}(1)$ and then the number of the sum is $\mathcal{O}(1)$.
So this part will give a leading-order contribution that is $\mathcal{O}(N)$.
\end{enumerate}

In summary, when we evaluate $P_2(w)$, we need to consider only
the $n=(L+1)/2+k$ with $k$ being $\mathcal{O}(1)$ part of the second finite sum
as a leading-order contribution.
We can evaluate \eqref{eq:77} and \eqref{eq:78} in a similar way,
and it is not difficult to conclude that the leading-order contribution
is only from $n=(L+1)/2+k$ of the finite sum part.

As for $\Theta(v,w) \Xi(v,w)$ in \eqref{eq:79},
one can carry out the same analysis as here.
In this case, there is no enhancement factor $1/(2n-L)$ in the sum,
and then the whole contribution turns out to be subleading compared
to the first $\Xi(v,w)$ term, with the $p^+ = \sqrt{N}$
factor taken into account.



 \end{document}